\documentclass[aps,prx,twocolumn,showpacs]{revtex4-1}
\usepackage{bm}
\usepackage{mathrsfs}
\usepackage{amsmath}
\usepackage{amssymb}
\usepackage{graphicx}
\usepackage{amsfonts}
\usepackage{amsthm}
\usepackage{color}
\usepackage{dcolumn}
\usepackage{txfonts}
\usepackage[caption=false]{subfig}
\begin{document}

\title{Cosmology emerging as the gauge structure of a nonlinear quantum system}
\author{Chon-Fai Kam}
\author{Ren-Bao Liu}
\email{Corresponding author. Email: rbliu@phy.cuhk.edu.hk}
\homepage{URL: http://www.phy.cuhk.edu.hk/rbliu}
\affiliation{Department of Physics, Centre for Quantum Coherence, and Institute of Theoretical Physics,
The Chinese University of Hong Kong, Shatin, New Territories, Hong Kong, China}

\begin{abstract}
Berry phases and gauge structures in parameter spaces of quantum systems are the foundation of a broad range of quantum effects such as quantum Hall effects and topological insulators. The gauge structures of interacting many-body systems, which often present exotic features, are particularly interesting. While quantum systems are intrinsically linear due to the superposition principle, nonlinear quantum mechanics can arise as an effective theory for interacting systems (such as condensates of interacting bosons). Here we show that gauge structures similar to curved spacetime can arise in nonlinear quantum systems where the superposition principle breaks down. In the canonical formalism of the nonlinear quantum mechanics, the geometric phases of quantum evolutions can be formulated as the classical geometric phases of a harmonic oscillator that represents the Bogoliubov excitations. We find that the classical geometric phase can be described by a de Sitter universe. The fundamental frequency of the harmonic oscillator plays the role of the cosmic scale factor and the classical geometric phase is an integral of a differential angle 2-form, which is half of the curvature 2-form of the associated de Sitter universe. While the gauge structure of a linear quantum system presents monopole singularity at energy level degeneracy points, nonlinear quantum systems, corresponding to their quantum critical surfaces in the parameter spaces, exhibits a conic singularity in their gauge structure, which mimics the casual singularity at the big bang of the de Sitter universe. This finding opens up a new approach to studying the gauge and topological  structures of interacting quantum systems and sets up a new stage for quantum simulation of fundamental physics.
\end{abstract}

\maketitle
\section{introduction}
Quantum phases are essential in many aspects of quantum physics, many-body physics, and quantum field theories. A spectacular feature of quantum phases is the appearance of geometric phases in an adiabatic process. In 1984, Berry discovered that in addition to the conventional dynamical phase, a geometric phase shift of a wave function is induced by a cyclic adiabatic change of parameters \cite{berry1984quantal}, which depends only on the shape of the cycle in the parameter space. Specifically, geometric phases arise from the overlap of coherent states \cite{Wen} along a closed path in the space of quantum states and play an indispensable role in the development of gauge field theories \cite{Wilczek}. Almost at the same time as Berry discovered Berry phases, in discussions of the relations between geometric phases and Chern integers \cite{Simon}, Simon recognized that the geometric phase is precisely the holonomy in fiber bundle theory \cite{Kobayashi} - while the wave function is single valued in the space of quantum states, it can be multi-valued around a cycle in the space of parameters. Following insights from Berry and Simon, Hannay \cite{Hannay, Berry85} showed that the shift of a classical phase angle in response to a cyclic adiabatic change of parameters is also a manifestation of the holonomy effect. It soon became clear that geometric phases reveal more than just phases. Geometric phases can be powerful tools for investigating a wide variety of intriguing properties of gauge field theories and are the basis of a broad range of phenomena and applications, such as quantum Hall effects \cite{QuantumHallEffect}, topological insulators and superconductors \cite{CKane, SCZhang}, artificial gauge fields in cold atomic gases \cite{ArtificialGaugePotentials}, holonomic quantum computation \cite{NonAbelianAnyons, nonAbelian} and quantum interference effects in single-molecule magnets \cite{MagneticMolecularClusters, MolecularSpintronics}. 

\begin{figure}
\begin{center}
\includegraphics[width=\columnwidth]{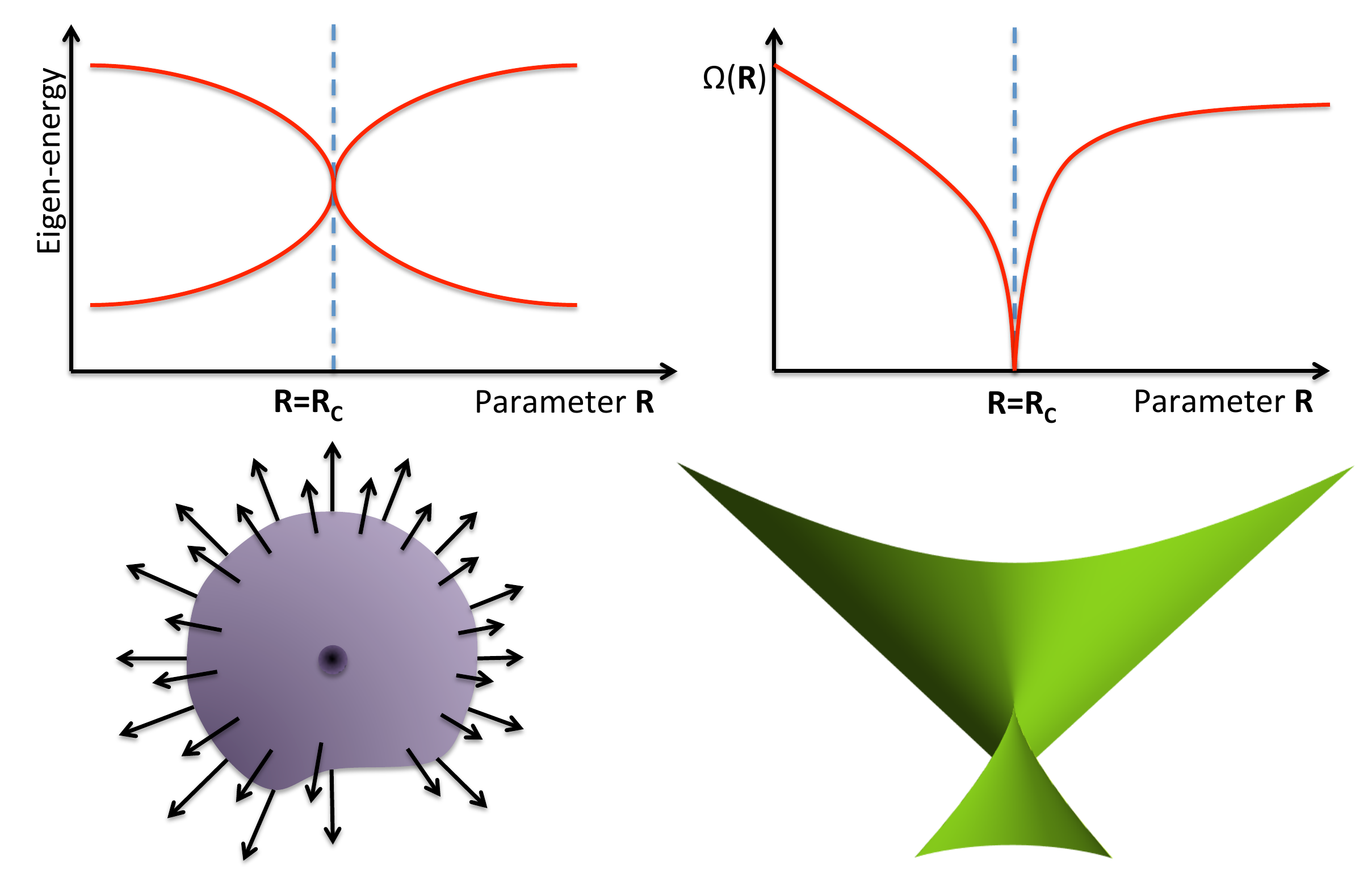}
\end{center}
\caption{The difference in the gauge structure anomalies between linear and nonlinear quantum systems. In linear quantum systems, the degeneracy of energy levels is associated with a magnetic monopole in the parameter space; by contrast, in nonlinear quantum systems, the vanishing of the fundamental frequency of the Bogoliubov excitations is associated with a critical surface in the parameter space. Here the critical surface is shown to have a swallowtail singularity at the origin.}
\label{fig:MonopoleTOSurface}
\end{figure}A fundamental aspect of geometric phases is the emergence of magnetic monopole singularity \cite{Dirac, TTWu} associated with the degeneracy of energy levels \cite{GaugeStructure, AharonovAnandan}. In a general evolution of quantum states, the geometric phase is characterized by a gauge invariant field, namely, Berry curvature. Berry curvature diverges at the degeneracy points, and can be regarded as an effective magnetic field with the point of degeneracy acting as its source, that is, a magnetic monopole in the parameter space (see Fig.\ref{fig:MonopoleTOSurface}). The existence of magnetic monopoles reflects the global nature of the parameter space and is invariant under local perturbations to the spectra \cite{EffectiveAction}.

\begin{figure}
\begin{center}
\includegraphics[width=\columnwidth]{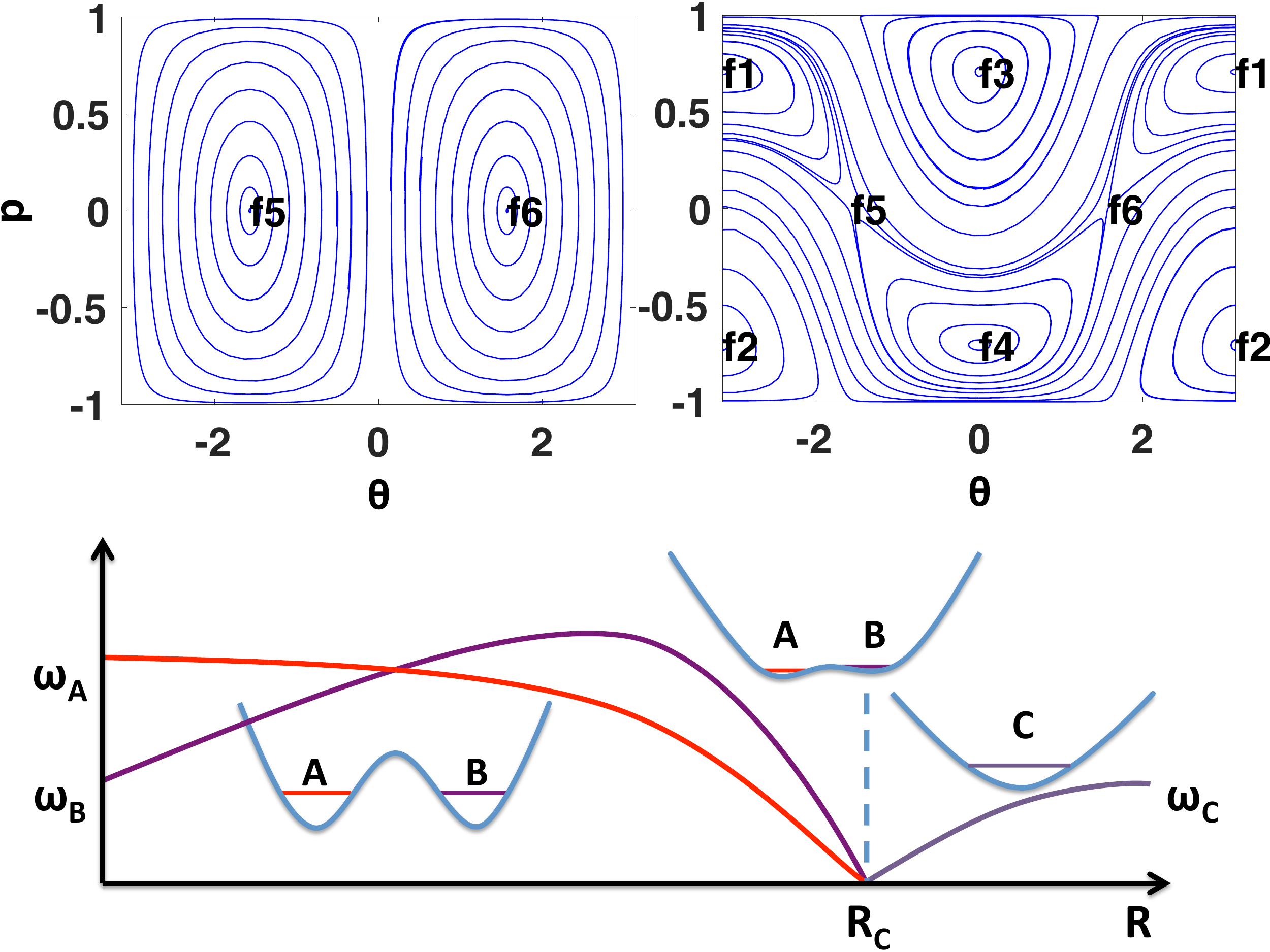}
\end{center}
\caption{Schematic of the Bogoliubov excitation spectrum of a nonlinear quantum system as a function of the control parameter $R$. Here $R_c$ is a critical value of $R$. For $R<R_c$, there are two Bogoliubov modes $A$ and $B$ with fundamental frequencies $\omega_A(R)$ and $\omega_B(R)$, respectively. The two fundamental frequencies drop to zero at $R=R_c$, which induces the mode softening. For $R>R_c$, the two modes merges into a new Bogoliubov mode $C$ with fundamental frequency $\omega_C(R)$. Here, the phase space portrait is obtained from the two mode-model in Eq.\eqref{NonlinearQuantumSystem} with $\Delta=\epsilon=0$, where the fixed points $f_5$ and $f_6$ correspond to $(\bar{p},\bar{\theta})=(0,\pm\pi/2)$. Surrounding the fixed points $f_5$ and $f_6$, there are two modes A and B with finite fundamental frequencies $\sqrt{\alpha\gamma-\beta^2}$, both the fundamental frequencies drop to zero at $\alpha\gamma-\beta^2=0$.}
\label{fig:Bogoliubov}
\end{figure}

Geometric quantum phases of interacting many-body systems are particularly interesting for their exotic features \cite{NiuQ, WuB, AnomalousMonopoles}. While quantum systems are intrinsically linear due to the superposition principle, nonlinear quantum mechanics can arise in interacting many-body quantum systems, such as condensates of interacting bosons \cite{BECBook} and quantum nanomagnets \cite{QuantumNanomagnets}. For example, the dynamics of the order parameter of interacting bosons can be effectively described by a nonlinear Schr\"{o}dinger equation \cite{Gross, LPPitaevskii}. In contrast with linear quantum systems, the superposition principle is no longer valid in nonlinear systems \cite{Weinberg}. Since quantum phases result essentially from the superposition principle, the definition of geometric phases of nonlinear quantum systems is subtle due to the breakdown of the linear superposition principle. Particularly, quantum phase transitions \cite{Sachdev} can occur in the nonlinear quantum systems with the change of external parameters. At the critical surfaces in the parameter spaces where the quantum phase transitions occur, the order parameter vanishes and so the adiabatic phases of the order parameter is expected to present singularity (see Fig.\ref{fig:MonopoleTOSurface}). The magnetic monopole paradigm is insufficient to reflect such singularity associated with the quantum critical phenomena \cite{SpinCriticality}. In nonlinear quantum systems, the central concept involved is the elementary Bogoliubov excitations \cite{NiuQ, WuB}, whose fundamental frequencies vary with external parameters. The fundamental frequencies of the Bogoliubov excitations vanish at the critical surfaces and the divergence of the time scale, that is, the inverse fundamental frequency at the critical surface indicates emergence of singularity in the adiabatic evolution in the parameter space (see Fig.\ref{fig:Bogoliubov}).

The canonical formalism of quantum mechanics has been introduced to formulate the geometric phases of nonlinear quantum systems so as to overcome the difficulties arising from the lack of superposition principle \cite{LBFu}. In this paper we will employ this formalism to explore connections between singularity of geometric phases and quantum critical phenomena. In the canonical formalism, the wave function is regarded as a classical field, which can be used to describe the order parameter of a Bose-Einstein condensate \cite{BECtunneling} or a collective spin system \cite{YuShi06}. The geometric phases of quantum evolutions are then formulated as the Hannay phases \cite{Hannay, Berry85} of classical harmonic oscillators that correspond to the Bogoliubov excitations of the quantum many-body systems \cite{NiuQ, WuB, LBFu}. For a discrete system with a mode index $k$, such as a Bose-Einstein condensate in a double-well trap, the time evolution of the wave function is governed by a set of coupled equations \cite{Weinberg}
\begin{equation}
i\frac{d\psi_k}{dt}=\frac{\partial}{\partial \psi^*_k}H(\psi,\psi^*,\mathbf{R}).
\end{equation}
Here $H$ is a real function of $\psi$ and $\psi^*$ depending on some external parameters $\mathbf{R}$. For instance, the coherent atomic tunneling between two Bose-Einstein condensates confined in a double-well trapping potential is described by the nonlinear Hamiltonian
\begin{equation*}
H=\epsilon\left(|\psi_1|^2-|\psi_2|^2\right)+\Delta\left(\psi_1^*\psi_2+\psi_1\psi_2^*\right)+\frac{\gamma}{2}\left(|\psi_1|^2-|\psi_2|^2\right)^2,
\end{equation*}
where $\epsilon$ is the difference of the single-mode energies, $\Delta$ is the Josephson tunneling rate and $\gamma$ is the nonlinear parameter proportional to the overlap of the spatial wave functions that are localized in each potential well. In the canonical formalism of the nonlinear quantum mechanics, the amplitude $p_k=|\psi_k|^2$ and the phase $\theta_k=\arg \psi_k$ of the wave amplitudes form a pair of canonical coordinates and the time evolution of the wave amplitude can be mapped to the corresponding classical dynamics. In this simple model, the classical Hamiltonian has the form \cite{NiuQ}
\begin{equation}
H=\epsilon p+ \frac{\gamma}{2}p^2+\Delta\sqrt{1-p^2}\cos\theta,
\end{equation}
where $p=p_1-p_2$ is the population imbalance between the two wells and $\theta=\theta_2-\theta_1$ is the relative phase of the two macroscopic wave functions. Notice that the global phase $\lambda=\theta_1+\theta_2$ is absent in the classical Hamiltonian, as a consequence of the conservation of total population $p_1+p_2$. For the case of a symmetric well, the Hamiltonian describes a classical nonrigid pendulum of tilt angle $\theta$ and a length proportional to $\sqrt{1-p^2}$.

In the canonical formalism of nonlinear quantum mechanics, the linearized dynamics near the fixed points are characterized by the fundamental frequencies of vibrations which correspond to the Bogoliubov excitations from the ground states. In the absence of nonlinearity, the fundamental frequencies are equivalent to the energy level spacing. However, the appearance of nonlinearity causes a bifurcation of the classical dynamics, which results in a qualitative change in the topology of the trajectories in the phase space. In particular, the bifurcation of the dynamics in the phase space implies the existence of quantum criticality in the original quantum system. As an example illustrated in Fig.\ref{fig:Bogoliubov}, the variation of the control parameters results in a change of the topological type of the dynamics. In the simplest case, the topological type only depends on a single control parameter $R$. For $R<R_c$, there are two distinct Bogoliubov modes $A$ and $B$ with different fundamental frequencies $\omega_A(R)$ and $\omega_B(R)$ respectively. At the critical point $R=R_c$, the two Bogoliubov modes experience mode softening with their fundamental frequencies dropping to zero. For $R>R_c$, the two discrete modes merge into a new Bogoliubov mode $C$. In this regard, the quantum criticality in a nonlinear quantum system is not induced by the degeneracy of the energy levels, but rather is caused by the softening of the Bogoliubov modes. Such an analysis can be applied to cases where there are more than one control parameters and there are various numbers of Bogoliubov modes before and after the phase transitions. When the control parameters are adiabatically varying, the disappearance of Bogoliubov modes and mergence of new ones cause the quantum criticality at a critical surface in the parameter space. Near the critical surface the system exhibits a mode softening, that is, the oscillation has an infinite period and all the energy levels collapse. In view of these facts, the oscillation period should be regarded as the clock of the system as it determines the characteristic time scale of the dynamics. This observation leads to a natural theoretic description of the geometric phase in the presence of quantum critical phenomena not based on traditional magnetic monopole paradigm, but in terms of the evolution of spacetime in classical relativity.

Here we report our discovery that the classical geometric phase of a generalized harmonic oscillator that correspond to the Bogoliubov mode of a nonlinear quantum system can be explained by the global geometry of a de Sitter universe \cite{Adler} described qualitatively by the Friedmann-Lema"tre-Robertson-Walker metric \cite{Misner}. In our method, the fundamental frequency of the oscillation near the critical surface plays the role of the cosmic scale factor \cite{Adler, Misner}, and the classical geometric phase is an integral of a differential 2-form that exhibits conic singularity similar to the casual singularity at the big bang \cite{Hawking} of the de Sitter universe (see Fig.\ref{fig:cone}).

\section{MODEL and Geometric phases}
As an example to demonstrate the quantum criticality beyond the paradigm of magnetic monopole singularity, we consider the Hamiltonian of a nonlinear quantum system with two wave amplitudes $\psi_1$ and $\psi_2$, which up to fourth order can be written as
\begin{align}\label{NonlinearQuantumSystem}
H&=\Delta\left(\psi_1^*\psi_2+\psi_1\psi_2^*\right)+\epsilon\left(|\psi_1|^2-|\psi_2|^2\right)+\frac{\alpha}{2}\left(\psi_1^*\psi_2+\psi_1\psi_2^*\right)^2\nonumber\\
&+\beta\left(\psi_1^*\psi_2+\psi_1\psi_2^*\right)\left(|\psi_1|^2-|\psi_2|^2\right)+\frac{\gamma}{2}\left(|\psi_1|^2-|\psi_2|^2\right)^2,
\end{align}
where $\Delta$, $\epsilon$, $\alpha$, $\beta$ and $\gamma$ are time-dependent external parameters that are characteristics of the system (see Appendix A for physical realization of the Hamiltonian in a double-well BEC). As the Hamiltonian is invariant under the global phase transformation, the total probability is conserved, $|\psi_1|^2+|\psi_2|^2=1$, and we can make the substitution $\psi_k=\sqrt{p_k}e^{i\theta_k}$, which yields $p_1+p_2=1$. If we define $p=p_1-p_2$ and $\theta=\theta_2-\theta_1$, the Hamiltonian becomes
\begin{equation*}
H=\epsilon p+\frac{\gamma}{2}p^2+(\Delta+\beta p)\sqrt{1-p^2}\cos \theta+\frac{\alpha}{2}(1-p^2)\cos^2\theta.
\end{equation*}
For the case of $\Delta=\epsilon=0$, $\bar{p}=0$ and $\bar{\theta}=\pi/2$ is a fixed point of the classical dynamics described by the canonical formalism of the nonlinear quantum system. Near the fixed point, the linearized Hamiltonian has the form of a generalized harmonic oscillator with time-dependent coefficients,
\begin{equation}
H\approx(\alpha(t)\theta^2+2\beta(t)p\theta+\gamma(t)p^2)/2.
\end{equation}
For $\alpha\gamma>\beta^2$, the Hamiltonian describes stable oscillations with elliptical trajectories in the phase plane; for $\alpha\gamma<\beta^2$, the origin is a saddle fixed point, and the trajectory contours become hyperbolae in the phase space. For a given energy $E$, the area of the trajectory ellipse is $2\pi E/(\alpha\gamma-\beta^2)^{1/2}$, a point on the ellipse is denoted by an angle variable $\Theta$ and the frequency of oscillation is $\omega=\sqrt{\alpha\gamma-\beta^2}$. The surface $\alpha\gamma=\beta^2$ defines a critical surface in the parameter space. 

Now we consider the geometric phase of the nonlinear quantum system. In considering our oscillator with slowly varying parameters, we notice that as the rate of change of the parameters approaches zero, the ratio of the energy $E$ to the frequency $\omega$ remains unchanged during the entire process so that the adiabatic condition is satisfied \cite{Arnold}. Hence, the action variable $I=E/\omega$ is an adiabatic invariant \cite{ArnoldMath} of the generalized harmonic oscillator. Therefore, the position of the oscillator on the ellipse when the Hamiltonian adiabatically evolves along a circuit $C$ in the parameter space after a long time $T$ is, $\Theta=\omega T+\Delta\Theta$, where $\Theta$ is the shift of the angle variable in response to the cyclic adiabatic change of parameters. In terms of the differential form, the classical adiabatic angle (Hannay phase) is an integral of the angle 2-form \cite{ArnoldMath}, $\Delta\Theta=\int_{\partial S=C} W$, where $S$ is an arbitrary open surface in the parameter space whose boundary is $C$. The angle 2-form for the generalized harmonic oscillator can be written explicitly \cite{Hannay, Berry85}
\begin{figure*}
	\subfloat[\label{sfig:cone}]{%
	\includegraphics[width=1.0\columnwidth]{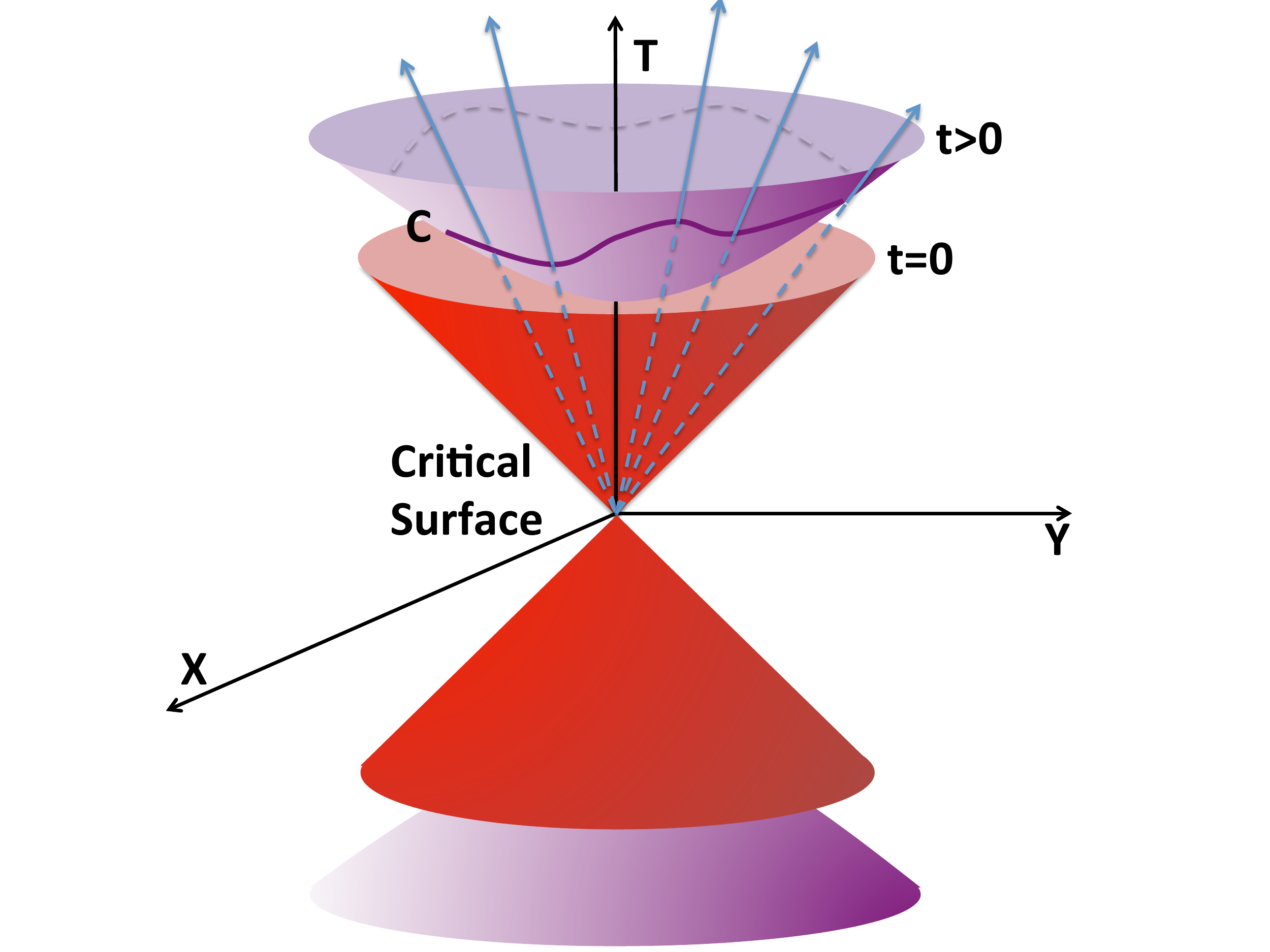}%
	}\hfill
	\subfloat[\label{sfig:penrose}]{%
 	 \includegraphics[width=1.0\columnwidth]{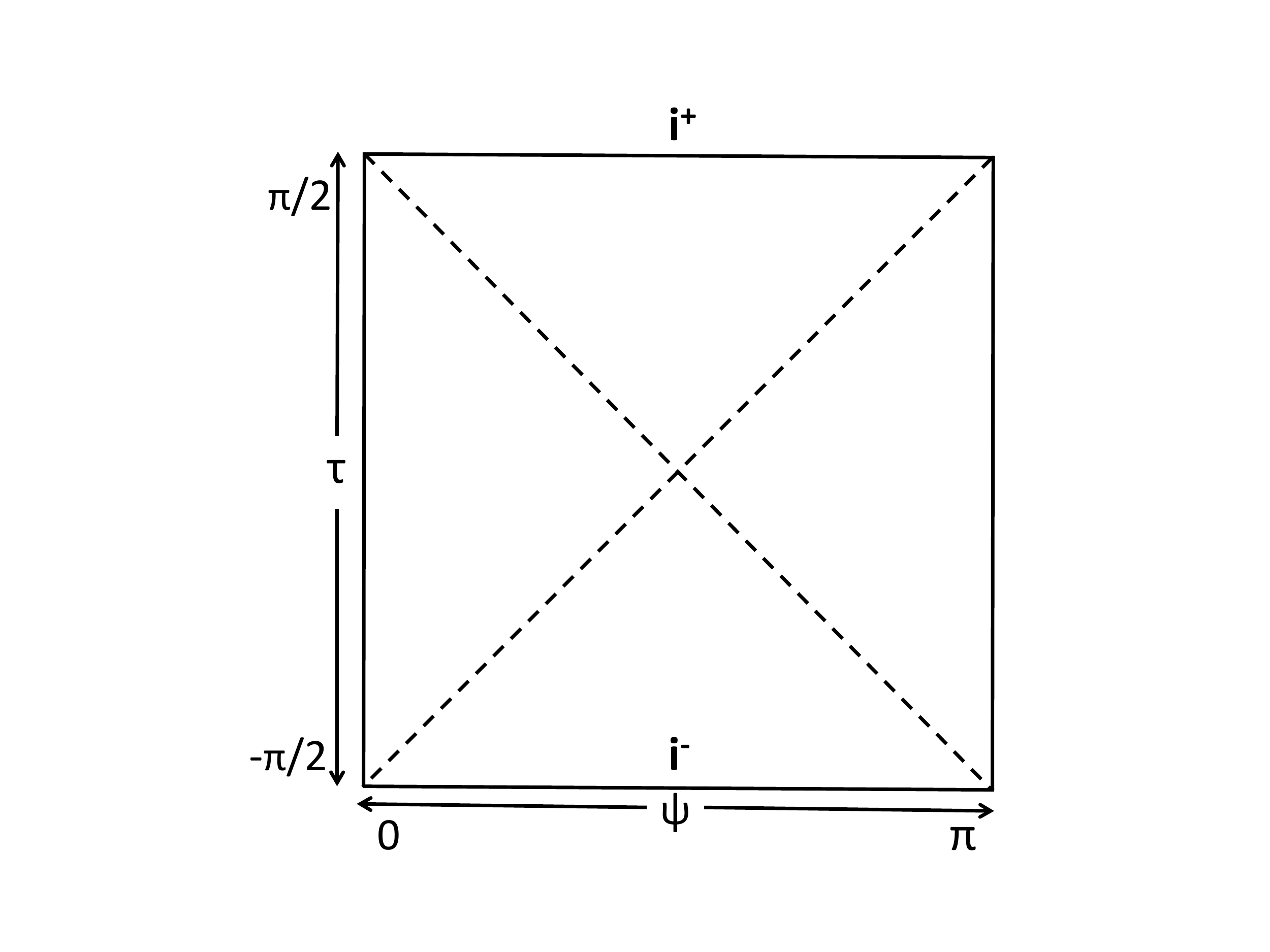}%
	}
\caption{Geometry of the classical adiabatic angle in the parameter space. (a). $T$, $X$ and $Y$ denote the external parameters of the Hamiltonian. The red cone locates the critical surface in the parameter space where the proper distance vanishes. $C$ denotes a closed curve and the classical adiabatic angle is the area bound by $C$ on the unit hyperboloid. (b). The Penrose diagram shows the causal structure of the de Sitter space, which is obtained by the usage of the conformal coordinates $(\tau,\psi, \phi)$ defined by $T=\tan\tau$, $r=\cos^{-1}\tau\sin\psi$ and $Z=\cos^{-1}\tau\cos\psi$, where $X=r\cos\phi$ and $Y=r\cos\phi$. In the conformal coordinates, the metric has the form $ds^2=\cos^{-2}\tau(-d\tau^2+d\psi^2+\sin^2\psi d\phi^2)$. Each point in the plot represents a circle and the left and right hand sides correspond to the north ($\psi=0$) and south ($\psi=\pi$) poles respectively. The upper and lower sides labeled $i^+$ and $i^-$ denote the infinite past and future respectively.}
\label{fig:cone}
\end{figure*}\begin{equation}
W=\frac{\alpha d\beta \wedge d\gamma+\beta d\gamma \wedge d\alpha+\gamma d\alpha \wedge d\beta}{4(\alpha\gamma-\beta^2)^{3/2}}.
\end{equation}
To better display the geometry, we introduce a change of variables $\alpha=T+X$, $\gamma=T-X$ and $\beta=Y$, such that
\begin{equation}
W=\frac{T dX\wedge dY+X dY \wedge dT+Y dT \wedge dX}{2(T^2-X^2-Y^2)^{3/2}}.
\end{equation}
If we denote $X^0=T$, $X^1=X$ and $X^2=Y$, the angle 2-form of the generalized harmonic oscillator is invariant under the local scale transformation $X'^{\mu}=\lambda(X^\mu)X^\mu$ and the \textit{SO}(2,1) transformation $X'^{\mu}=\Lambda^\mu_\nu X^\nu$, where $\lambda(X^\mu)$ is a local scale factor and $\Lambda^\mu_\nu$ is an element of the Lorentz group. An important feature is that when the oscillation frequency $\omega=\sqrt{T^2-X^2-Y^2}$ approaches zero, the angle 2-form exhibits a conical singularity in the parameter space (see Fig.\ref{sfig:cone}). In other words, as the parameters get infinitesimally close to the cone $T^2-X^2-Y^2=0$, the classical adiabatic angle diverges. 

The conic singularity in the parameter space is similar to the causal singularity of the light cone in the Minkowski space. Such an analogue implies that the external parameters may be identified with the coordinates in the Minkowski space with adequate constraints. A natural assumption is that the spacetime is homogeneous and isotropic, then we can define a cosmic clock by the oscillation frequency. We will demonstrate that the 2+1 dimensional de Sitter space is an adequate choice. 

\section{cosmology analogue}
In this section, we will analyze the relation between the classical adiabatic angle and the cosmology in 2+1 dimensions in details. The 2+1 dimensional de Sitter space is defined as the set of all points $(T,X,Y,Z)$ in the 3+1 dimensional Minkowski space subjected to the constraint \cite{Adler, Zee}
\begin{equation}
-T^2+X^2+Y^2+Z^2=1,
\end{equation}
where the metric of the Minkowski space is given by
\begin{equation}
ds^2=-dT^2+dX^2+dY^2+dZ^2.
\end{equation}
The de Sitter space is maximally symmetric \cite{WeinbergBook}, and thus the Riemann curvature tensor is fully determined by the metric tensor, $R_{\mu\nu\lambda\sigma}=g_{\mu\lambda}g_{\nu\sigma}-g_{\mu\sigma}g_{\nu\lambda}$. It is straightforward to check that the de Sitter space is a solution of Einstein field equations with a positive cosmological constant $\Lambda=1$. After contraction of indices, we obtain the Ricci tensor $R_{\mu\nu}=2g_{\mu\nu}$ and the scalar curvature $R=6$. Among the various expressions of the de Sitter space, the coordinate choice $Z=\cosh t$, $T=\sinh t\tilde{T}$, $X=\sinh t\tilde{X}$ and $Y=\sinh t\tilde{Y}$ yields $ds^2=-dt^2+\sinh^2td\sigma^2$, where $d\sigma^2=-d\tilde{T}^2+d\tilde{X}^2+d\tilde{Y}^2$ is the spatial metric with coordinates $\tilde{T}$, $\tilde{X}$ and $\tilde{Y}$ satisfying $\tilde{T}^2-\tilde{X}^2-\tilde{Y}^2=1$, which describes the two-dimensional unit hyperboloid. In the hyperbolic coordinates $\tilde{T}=\cosh\psi$, $\tilde{X}=\sinh\psi\cos\phi$ and $\tilde{Y}=\sinh\psi\sin\phi$, we obtain the standard metric on the unit hyperboloid, $d\sigma^2=d\psi^2+\sinh^2\psi d\phi^2$, from which we see that the volume form of the unit hyperboloid is $\sinh\psi d\psi\wedge d\phi$. Therefore, the metric of the 2+1 dimensional de Sitter space in the coordinates $(t,\psi,\phi)$ is precisely the Friedmann-Lema"tre-Robertson-Walker metric \cite{Misner}, which describes a homogeneous and isotropic expanding universe in 2+1 dimensions \cite{Adler, Misner, Hawking}
\begin{equation}
ds^2=-dt^2+a^2(t)(d\psi^2+\sinh^2\psi d\phi^2),
\end{equation}
where $t\in(-\infty,\infty)$ is the coordinate time, $\psi\in(0,\infty)$ is the hyperbolic angle, $\phi\in(0,2\pi)$ is the circular angle and $a(t)=\sinh t$ is the cosmic scale factor \cite{Misner} of the de Sitter universe. At a given time, the universe corresponds to a slice of the de Sitter hyperboloid at a fixed $Z$. At $t=0$, the spacetime degenerates into a single point $(T,X,Y,Z)=(0,0,0,1)$, which corresponds to the big bang of the de Sitter universe \cite{Misner}. 

In the tetrad formalism of general relativity \cite{Weinberg, Wald, Carmeli}, the vielbein, the connection form and the curvature form are the basic quantities. The vielbeins for the two-dimensional unit hyperboloid at a given time are $e^1=a(t)d\psi$ and $e^2=a(t)\sinh\psi d\phi$. As the infinitesimal rotations of the vielbeins are described by the first Cartan structure equation $de^a+\omega^a_b\wedge e^b=0$ \cite{Kobayashi}, the only non-vanishing connection 1-form on the unit hyperboloid is $\omega_1^2=-\omega_2^1=\cosh\psi d\phi$. The curvature 2-form is obtained from the first Cartan structure equation $R_b^a=d\omega_b^a+\omega_c^a\wedge\omega_c^c$, and the only non-vanishing curvature 2-form on the unit hyperboloid is $R_1^2=-R_2^1=\sinh\phi d\psi \wedge d\phi$. Evidently, it is equal to the volume form of the unit hyperboloid. In terms of the coordinates in the Minkowski space, the curvature 2-form can be written as
\begin{equation}
R_1^2=\frac{TdX\wedge dY+XdY\wedge dT+YdT\wedge dX}{(T^2-X^2-Y^2)^{3/2}},
\end{equation}
where $T$, $X$ and $Y$ satisfy $\omega^2=T^2-X^2-Y^2=a^2(t)$. With the fundamental frequency of oscillation mapped to the cosmic scale factor of the expanding universe, the conical singularity of the angle 2-form in the parameter space of the harmonic oscillator can be understood as the causal singularity of the de Sitter universe at the beginning of time (see Fig.\ref{fig:cone}). As a result, we obtain an important relation between the classical adiabatic angle of the generalized harmonic oscillator and the 2+1 dimensional de Sitter universe: the angle 2-form is half of the curvature 2-form of the associated de Sitter universe
\begin{equation}
W=\frac{1}{2}R_1^2,
\end{equation}
 such that the classical adiabatic angle for any surface $S$ bounded by $C$ on the unit hyperboloid is
\begin{equation}
\Delta\Theta=\int_{\partial S=C}W=\frac{1}{2}\int_0^{2\pi}\int_0^{\psi(\phi)}\sinh\psi d\psi d\phi=\frac{1}{2}A(C),
\end{equation}
where $A(C)$ is the area on the unit hyperboloid subtended by $C$ at the origin. As shown in Fig.\ref{fig:cone}, an arbitrary circuit in the parameter space, now recognized as the 2+1 dimensional de Sitter universe, can be projected into the two-dimensional unit hyperboloid, and thus the resulting classical adiabatic angle is half of the solid angle subtended by the circuit at the origin. Moreover, the critical surface as a single point $(T,X,Y,Z)=(0,0,0,1)$ can be understood as the boundary of the Penrose diagram for the associated de Sitter space, where the proper distance vanishes.

\section{physical realization}
Here we outline an experimental proposal to realize the classical geometric phase based on a Bose-Einstein Condensate (BEC) in an asymmetric double-well potential. For further details, see Appendix A.

In this proposal, the BEC in an double-well potential is described by two weakly coupled macroscopic wave functions separated by a potential barrier \cite{Pitaevskii}. Denoting the two wave amplitudes by $\psi_1$ and $\psi_2$, the Hamiltonian has the form
\begin{align}
H&=\epsilon_1|\psi_1|^2+\epsilon_2|\psi_2|^2+\frac{U_1}{2}|\psi_1|^4+\frac{U_2}{2}|\psi_2|^4\nonumber\\
&+(K+U_{12}|\psi_1|^2+U_{21}|\psi_2|^2)(\psi_1^*\psi_2+\psi_1\psi_2^*)\nonumber\\
&+2I|\psi_1|^2|\psi_2|^2+\frac{I}{2}(\psi_1^{*2}\psi_2^2+\psi_1^2\psi_2^{*2}),
\end{align}
where $\epsilon_1$ and $\epsilon_2$ are the single-mode energies, $U_1$ and $U_2$ are the on-site interaction energies, $K+U_{12}|\psi_1|^2+U_{21}|\psi_2|^2$ is the renormalized tunneling rate that depends on the populations of the two condensates, and the last two terms in $I$ are the inter-well interaction and the inter-well pair tunneling respectively. The dynamics of the system is governed by only two variables, the fractional population imbalance $p=(N_1-N_2)/(N_1+N_2)$ and the quantum relative phase $\theta=\theta_2-\theta_1$ between the left and right condensates. The resulting quantum dynamics in an asymmetric double-well potential is described by
\begin{equation}
\dot{p}=E_J\sqrt{1-p^2}\sin\theta, \dot{\theta}=E_C-E_J\frac{ p\cos\theta}{\sqrt{1-p^2}},
\end{equation}
where $E_J=\Delta+\beta p+\alpha\sqrt{1-p^2}\cos\theta$ is the effective Josephson tunneling energy, $E_C=\epsilon+\gamma p+\beta\sqrt{1-p^2}\cos\theta$ is the effective energy difference between the two condensates. Here the coefficients $\Delta$, $\epsilon$, $\alpha$, $\beta$ and $\gamma$ are given by $\Delta=2K+U_{12}+U_{21}$, $\epsilon=\epsilon_1-\epsilon_2+(U_1-U_2)/2$, $\alpha=2I$, $\beta=U_{12}-U_{21}$ and $\gamma=(U_1+U_2)/2-I$, where $\Delta$ is the static tunneling energy, $\epsilon$ is the difference of the single-mode energies and $\alpha$, $\beta$ and $\gamma$ are determined by the overlap of the spatial wave functions that are localized in each well. The Josephson tunneling energy $E_J$, which explicitly includes the nonlinear interaction effects, depends significantly on the values of $\alpha$ and $\beta$, and in turn on the inter-well pair tunneling rate and the difference between the interaction-assisted tunneling energies.

As expected from the Josephson effect, the population imbalance and the relative phase execute harmonic oscillations surrounding the fixed points of the dynamics. However, unlike the standard two-mode model, the system has novel fixed points for $E_J=E_C=0$, which are solved by
\begin{equation}
\bar{p}=\frac{\beta\Delta-\alpha\epsilon}{\alpha\gamma-\beta^2},\sqrt{1-\bar{p}^2}\cos\bar{\theta}=\frac{\beta\epsilon-\gamma\Delta}{\alpha\gamma-\beta^2}.
\end{equation}
Specifically, for $\epsilon=\Delta=0$, the mean values of the population imbalance and the relative phase are $\bar{p}=0$ and $\bar{\theta}=\pm\pi/2$, where the harmonic oscillations around the mean values are governed by $\dot{p}=\pm\beta p-\alpha\theta$ and $\dot{\theta}=\gamma p\mp \beta\theta$, which can be derived from the Hamiltonian of a generalized harmonic oscillator, $H=(\alpha\theta^2\mp 2\beta p\theta+\gamma p^2)/2$. This implies that an initial population imbalance and a small derivation of the $\pi/2$ phase induces a sinusoidal Josephson oscillations with a finite oscillation frequency $\omega=\sqrt{\alpha\gamma-\beta^2}$. In contrast, the Josephson oscillations of an initial population imbalance and a small relative phase for a symmetric double-well potential are governed by $\dot{p}=(\Delta+\alpha)\theta$ and $\dot{\theta}=-(\Delta+\alpha-\gamma)p$, which only describes an ordinary harmonic oscillator, as $\dot{p}$ is linearly proportional to $\theta$ but not a linear combination of $p$ and $\theta$.

In a realistic experiment, the BEC after initial evaporative cooling is loaded into an optical effective double well trap, which is created by the superposition of a periodic potential with a harmonic trapping potential \cite{BECtunneling, BECJosephson}
\begin{equation}
V=\frac{m}{2}\left(\omega_x^2(x-\Delta x)^2+\omega_y^2y^2+\omega_z^2z^2\right)+V_0\cos^2\left(\frac{\pi x}{d}\right),
\end{equation}
where $\omega_x$, $\omega_y$ and $\omega_z$ are the harmonic trapping frequencies, $d$ is the periodicity, $V_0$ is the potential depth and $\Delta x$ is the relative position shift of the two potentials. The initial population imbalance of the two wells is obtained by loading the condensate into an asymmetric double-well potential, which is created by a nonzero shift of the harmonic confinement with respect to the periodic potential. The parameters $\alpha$, $\beta$ and $\gamma$ can be tuned slowly by adjusting the potential barrier, the harmonic trapping frequencies and the relative shift of the two potentials independently \cite{AtomSQUID, asymmetricBEC}.

\section{summary}
Consideration of the nonlinear quantum system with three external parameters leads to a gauge field that exhibits a conic singularity similar to the casual singularity at the big bang of the 2+1 dimensional de Sitter universe. The gauge field is half of the curvature form of the de Sitter universe. A physical realization based on a BEC in an asymmetric double-well potential is proposed. This finding could be extended to higher dimensional spacetime in future studies when the nonlinear quantum system has more than three parameters.

\begin{acknowledgements}
This work was supported by Hong Kong RGC and CUHK VC's One-off Discretionary Fund.
\end{acknowledgements}

\begin{appendix}
\section{Two-mode model for asymmetric double well Bose-Einstein Condensates}
The many-body Hamiltonian that describes interacting bosons confined by an external potential $V_{ext}(\mathbf{r})$ is
\begin{align}
\hat{H}&=\int d\mathbf{r} \hat{\Psi}^\dagger(\mathbf{r})\left[-\frac{\hbar^2\nabla^2}{2m}+V_{ext}(\mathbf{r})\right]\hat{\Psi}(\mathbf{r})\nonumber\\
&+\frac{1}{2}\int d\mathbf{r} d\mathbf{r}' \hat{\Psi}^\dagger(\mathbf{r})\Psi^\dagger(\mathbf{r}')V(\mathbf{r}-\mathbf{r}')\hat{\Psi}(\mathbf{r}')\hat{\Psi}(\mathbf{r}),
\end{align}
where $\hat{\Psi}(\mathbf{r})$ and $\hat{\Psi}^\dagger(\mathbf{r})$ are the boson field operators that annihilate and create a particle at $\mathbf{r}$ respectively and $V(\mathbf{r}-\mathbf{r}')$ is the two-body interatomic potential \cite{Pitaevskii}. In the Heisenberg representation for the field operators, the time evolution of the field operator is determined by the Heisenberg equation
\begin{align*}
&i\frac{\partial \hat{\Psi}(\mathbf{r},t)}{\partial t}=[\hat{\Psi},\hat{H}]\\\nonumber
&=\left[-\frac{\hbar^2\nabla^2}{2m}
+V_{ext}(\mathbf{r})+\int d\mathbf{r}'\hat{\Psi}^\dagger(\mathbf{r}',t)V(\mathbf{r}-\mathbf{r}')\hat{\Psi}(\mathbf{r}',t)\right]\hat{\Psi}(\mathbf{r},t).
\end{align*}
In a dilute ultracold atomic gas, only the elastic binary collisions between individual atoms are relevant. The binary collisions are characterized by a single $s$-wave scattering length $a$, which is irrelevant to the expressions of the two-body potential. Hence, we can replace the two-body potential $V(\mathbf{r}-\mathbf{r}')$ with an effective interaction $g\delta(\mathbf{r}-\mathbf{r}')$, which results in 
\begin{equation*}
i\hbar\frac{\partial \hat{\Psi}(\mathbf{r},t)}{\partial t}=\left[-\frac{\hbar^2\nabla^2}{2m}
+V_{ext}(\mathbf{r})+g\hat{\Psi}^\dagger(\mathbf{r},t)\hat{\Psi}(\mathbf{r},t)\right]\hat{\Psi}(\mathbf{r},t),
\end{equation*}
where the coupling constant $g$ is related to the scattering length through $g=4\pi \hbar^2 a/m$. When BEC occurs, we can replace the field operator $\hat{\Psi}(\mathbf{r},t)$ with its mean-field value $\Phi(\mathbf{r},t)\equiv \langle \hat{\Psi}({\mathbf r})\rangle$ and obtain the time-dependent Gross-Pitaevskii equation for the condensate wave function \cite{Leggett}
\begin{equation}\label{GPEquation}
i\hbar\frac{\partial \Phi(\mathbf{r},t)}{\partial t}
=\left(-\frac{\hbar^2\nabla^2}{2m}+V_{ext}(\mathbf{r})+g|\Phi(\mathbf{r},t)|^2\right)\Phi(\mathbf{r},t).
\end{equation}
Here $\int |\Phi(\mathbf{r},t)|^2d\mathbf{r}=N$ is the number of condensed atoms. In a double-well potential, the Bose-Einstein condensate wave function $\Phi(\mathbf{r},t)$ can be written as a superposition of two time-independent spatial wave functions $\phi_1(\mathbf{r})$ and $\phi_2(\mathbf{r})$ that are localized in each well
\begin{equation}\label{TwoModeAnsatz}
\Phi(\mathbf{r},t)=\sqrt{N}[\psi_1(t)\phi_1(\mathbf{r})+\psi_2(t)\phi_2(\mathbf{r})],
\end{equation}
where $\psi_1(t)$ and $\psi_2(t)$ are the time-dependent modal amplitudes. The condensate wave functions in the two wells $\phi_1(\mathbf{r})$ and $\phi_2(\mathbf{r})$ are assumed to be real valued functions satisfying the orthonormal condition
\begin{equation}\label{NormalizedCondition}
\int \phi_i(\mathbf{r})\phi_j(\mathbf{r}) d\mathbf{r}=\delta_{ij}.
\end{equation}
Hence $|\psi_1|^2$ and $|\psi_2|^2$ represent the occupation probabilities for the two modes, and the normalization condition for the condensate wave function $\int |\Phi(\mathbf{r},t)|^2d\mathbf{r}=N$ leads to the conservation of the occupation probabilities, $|\psi_1|^2+|\psi_2|^2=1$. Substitution of Eqs.\eqref{TwoModeAnsatz} and \eqref{NormalizedCondition} into Eq.\eqref{GPEquation} yields immediately
\begin{subequations}\label{EquationOfMotion}
\begin{align}
i\dot{\psi_1}
&=\epsilon_1\psi_1 + K\psi_2+U_1|\psi_1|^2\psi_1+U_{12}(\psi_1^2\psi_2^*+2|\psi_1|^2\psi_2)\nonumber\\
&+I(2\psi_1|\psi_2|^2+\psi_1^*\psi_2^2)+U_{21}|\psi_2|^2\psi_2,\\
i\dot{\psi_2}
&=K\psi_1+\epsilon_2\psi_2+U_{12}|\psi_1|^2\psi_1+I(\psi_1^2\psi_2^*+2|\psi_1|^2\psi_2)\nonumber\\
&+U_{21}(2\psi_1|\psi_2|^2+\psi_1^*\psi_2^2)+U_2|\psi_2|^2\psi_2,
\end{align}
\end{subequations}
where the parameters $\epsilon_i$, $K$, $U_i$, $U_{ij}$ and $I$ are given by the following overlap integrals
\begin{subequations}
\begin{align}
\epsilon_i&=\int \phi_i(\mathbf{r})\left(-\frac{\hbar^2\nabla^2}{2m}+V_{ext}(\mathbf{r})\right)\phi_i(\mathbf{r}) d\mathbf{r},\\
K&=\int \phi_1(\mathbf{r})\left(-\frac{\hbar^2\nabla^2}{2m}+V_{ext}(\mathbf{r})\right)\phi_2(\mathbf{r}) d\mathbf{r},\\
U_i&=gN\int\phi_i^4(\mathbf{r})d\mathbf{r},\\
U_{ij}&=gN\int\phi_i^3(\mathbf{r})\phi_j(\mathbf{r})d\mathbf{r},\\
I&=gN\int\phi_1^2(\mathbf{r})\phi_2^2(\mathbf{r})d\mathbf{r}.
\end{align}
\end{subequations}
Here $\epsilon_i$ are the single-mode energies, $K$ is the tunneling rate of atoms between the two wells, and $U_i$ are the on-site interaction energies. These parameters are the same as those defined for the standard two-mode model \cite{AtomicTunneling}. The remaining parameters $U_{12}$, $U_{21}$ and $I$ include all the mixed terms in the spatial wave functions, and thus they are present only when the spatial wave functions have small but non-zero density on the other side. They were first introduced by Ananikian and Bergeman to include a renormalized tunneling rate to provide better agreement with numerical simulations and experimental results \cite{ITwoModeModel}. To show this, we rewrite Eqs.\eqref{EquationOfMotion} into the canonical formalism as $i\dot{\psi_k}=\partial H/\partial \psi_k^*$, where the classical Hamiltonian is
\begin{align}
H&=\epsilon_1|\psi_1|^2+\epsilon_2|\psi_2|^2+\frac{U_1}{2}|\psi_1|^4+\frac{U_2}{2}|\psi_2|^4\nonumber\\
&+(K+U_{12}|\psi_1|^2+U_{21}|\psi_2|^2)(\psi_1^*\psi_2+\psi_1\psi_2^*)\nonumber\\
&+2I|\psi_1|^2|\psi_2|^2+\frac{I}{2}(\psi_1^{*2}\psi_2^2+\psi_1^2\psi_2^{*2}).
\end{align}
The physical meanings of the parameters $U_{12}$, $U_{21}$ and $I$ can be understood from the Hamiltonian, where the $U_{12}|\psi_1|^2$ term contributes an interaction-assisted tunneling to the Hamiltonian and similarly for the $U_{21}|\psi_2|^2$ term. The last two terms in the Hamiltonian have different origins, where the first term $2I|\psi_1|^2|\psi_2|^2$ represents the inter-well interaction and the second term $I(\psi_1^{*2}\psi_2^2+\psi_1^2\psi_2^{*2})/2$ is the pair tunneling energy.

Quantitative analysis of the population variations of the two condensates and the macroscopic tunneling effects can be performed by using the canonical formalism. Let us make the substitutions $\psi_k=\sqrt{p_k}e^{i\theta_k}$, where $p_1$ and $p_2$ are the fractional populations of the Bose atoms at the two wells and $\theta_1$ and $\theta_2$ are the phases on the two sides of the barrier. If we define $\theta=\theta_2-\theta_1$, Eqs.\eqref{EquationOfMotion} can be written as
\begin{subequations}\label{Josephoson}
\begin{align}
\dot{p}_1&=+2K'\sqrt{p_1p_2}\sin\theta+2Ip_1p_2\sin2\theta,\\
\dot{p}_2&=-2K'\sqrt{p_1p_2}\sin\theta-2Ip_1p_2\sin2\theta, \\
\dot{\theta}_1&=-\epsilon'_1-K'_1\sqrt{\frac{p_2}{p_1}}\cos\theta-2Ip_2\cos^2\theta, \\
\dot{\theta}_2&=-\epsilon'_2-K'_2\sqrt{\frac{p_1}{p_2}}\cos\theta-2Ip_1\cos^2\theta.
\end{align}
\end{subequations}
where $\epsilon'_1=\epsilon_1+U_1p_1+Ip_2$, $\epsilon'_2=\epsilon_2+U_2p_2+Ip_1$ are the single-mode energies modified by the nonlinear interactions, $K'=K+U_{12}p_1+U_{21}p_2$ is the tunneling energy modified by the overlap of the spatial wave functions, and $K'_1=K'+2U_{12}p_1$ and $K'_2=K'+2U_{21}p_2$. The first pair of equations implies $\dot{p}_1=-\dot{p}_2$, which comes from the conservation of populations, $p_1+p_2=1$. The atomic current cross the barrier is $N\dot{p}_1$, or $-N\dot{p}_2$. When $p_1$ and $p_2$ are almost the same, the atomic current would be given by $J=J_0\sin\theta+I_0\sin 2\theta$, where $J_0=NK'$ and $I_0=NI/2$, and the phase evolution is determined by $\dot{\theta}=\epsilon_1-\epsilon_2+(K_1'-K_2')\cos\theta$. For the case when $I=0$ and $U_{12}=U_{21}$, we recover the Josephson equations, $J=J_0\sin\theta$ and $\dot{\theta}=\epsilon_1-\epsilon_2$ \cite{JosephsonEffects}. If we define $p=p_1-p_2$, the coupled-mode equations become
\begin{subequations}\label{GeneralTunnelingEquations}
\begin{align}
\dot{p}&=(\Delta+\beta p)\sqrt{1-p^2}\sin\theta+I(1-p^2)\sin 2\theta,\\
\dot{\theta}&=\epsilon+\gamma p+\frac{\beta(1-2p^2)-\Delta p}{\sqrt{1-p^2}}\cos\theta-\alpha p\cos^2\theta,
\end{align}
\end{subequations}
which can be derived from the Hamiltonian \cite{ModeExchangeBEC, asymmetricBEC}
\begin{equation*}
H=\epsilon p+\frac{\gamma}{2}p^2+(\Delta+\beta p)\sqrt{1-p^2}\cos\theta+\frac{\alpha}{2}(1-p^2)\cos^2\theta.
\end{equation*}
Here the coefficients $\Delta$, $\epsilon$, $\alpha$, $\beta$ and $\gamma$ are given by $\Delta=2K+U_{12}+U_{21}$, $\epsilon=\epsilon_1-\epsilon_2+(U_1-U_2)/2$, $\alpha=2I$, $\beta=U_{12}-U_{21}$ and $\gamma=(U_1+U_2)/2-I$. Note that the nonlinear interactions produce a temporal change in the tunneling energy, and the tunneling energy $\Delta+\beta p$ is proportional to the population imbalance. 

In a symmetric double well, we expect $\epsilon_1=\epsilon_2$, $U_1=U_2$ and $U_{12}=U_{21}$, which implies $\epsilon=\beta=0$. Then the time evolution for $p$ and $\theta$ are governed by
\begin{figure*}
\subfloat[$\epsilon=\Delta=0$, $\alpha=\gamma=1$, $\beta=0.5$\label{sfig:Beta05}]{%
  \includegraphics[width=1.0\columnwidth]{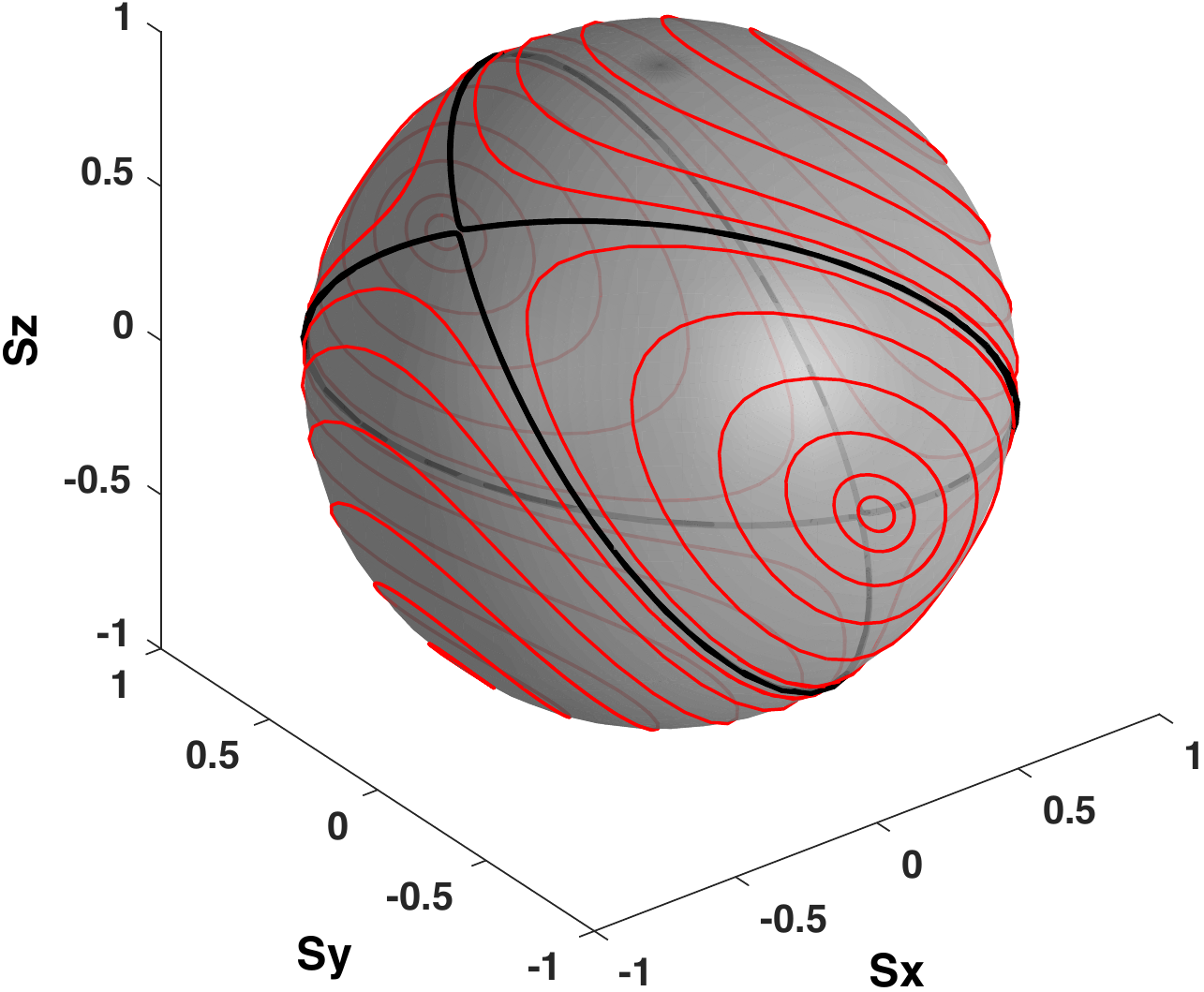}%
}\hfill
\subfloat[$\epsilon=\Delta=0$, $\alpha=\gamma=1$, $\beta=1.5$\label{sfig:testa}]{%
  \includegraphics[width=1.0\columnwidth]{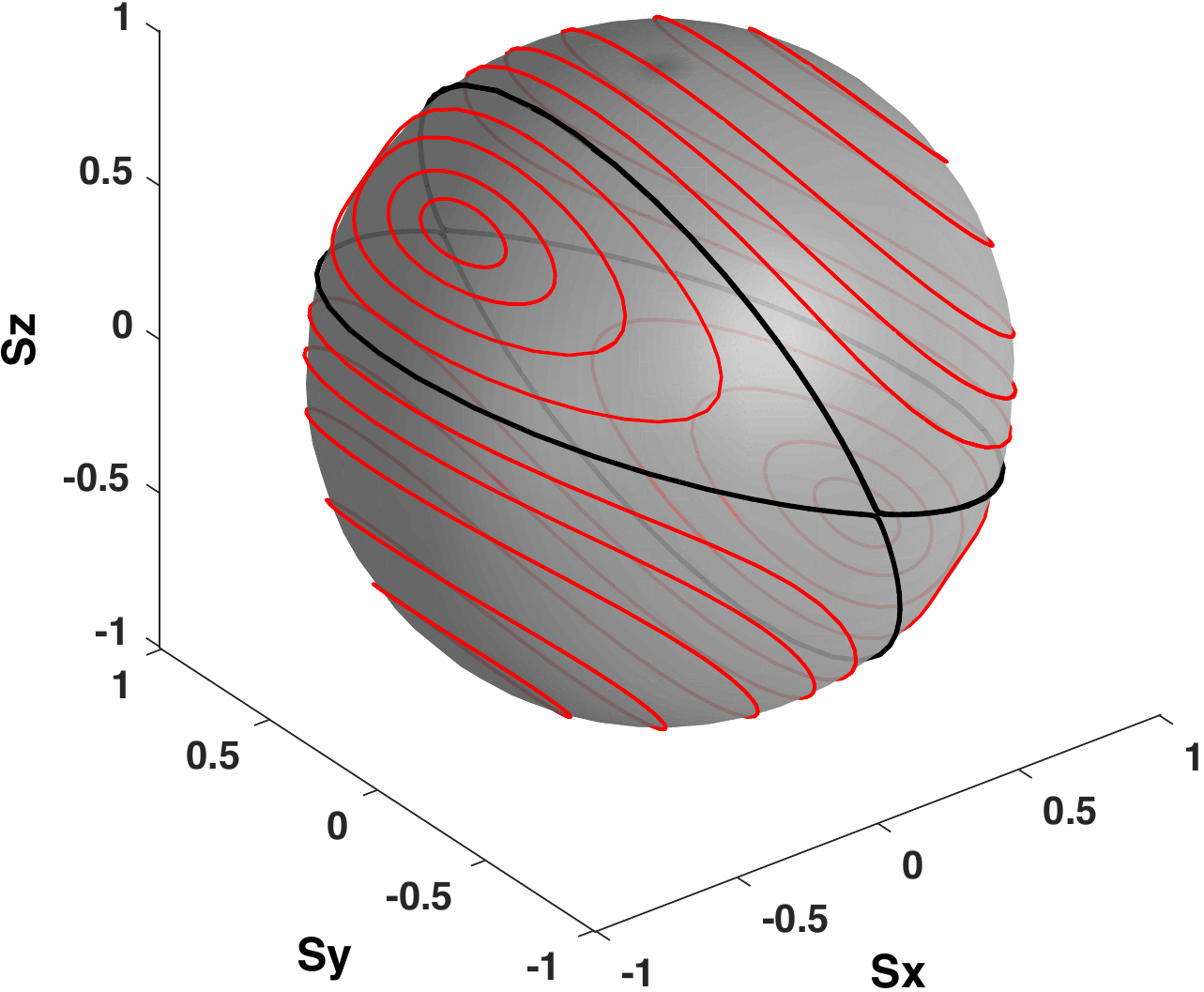}%
}
\caption{Classical spin representation of fixed points and Bogoliubov excitations of a double-well condensate for the symmetric case $(\epsilon=\Delta=0)$. The red curves show the phase trajectories for different energies $E=(\alpha S_x^2+2\beta S_xS_z+\gamma S_z^2)/2$. The bold black lines indicate the separatrices which divide the sphere into four distinct regions. The left and right panels have the same values of $\alpha$ and $\gamma$ ($\alpha=\gamma=1$) and the six fixed points are located at $(\bar{S}_x,\bar{S}_y,\bar{S}_z)=(0,\pm 1,0)$ and $(\pm 1/\sqrt{2}, 0, \pm 1/\sqrt{2})$ respectively. The left panel ($\beta=0.5$) shows the phase diagram for $\alpha\gamma-\beta^2>0$ in which the fixed points $(\bar{S}_x,\bar{S}_y,\bar{S}_z)=(0,\pm 1,0)$ are stable and the right panel ($\beta=1.5$) shows the phase diagram for $\alpha\gamma-\beta^2<0$ in which the points $(\bar{S}_x,\bar{S}_y,\bar{S}_z)=(0,\pm 1,0)$ become unstable.}
\label{SpinDynamics}
\end{figure*}
\begin{subequations}\label{SymmetricTunneling}
\begin{align}
\dot{p}&=\Delta\sqrt{1-p^2}\sin\theta+I(1-p^2)\sin 2\theta,\\
\dot{\theta}&=\gamma p-\frac{\Delta p}{\sqrt{1-p^2}}\cos\theta-\alpha p\cos^2\theta.
\end{align}
\end{subequations}
For $\bar{\theta}=0$ or $\pi$, the fixed points of Eqs.\eqref{SymmetricTunneling} are $\bar{p}=0$ or $\bar{p}=\pm\sqrt{1-\Lambda^2}$, where $\Lambda=\Delta/(\gamma-\alpha)$. In contrast, for the case when $\bar{\theta}\neq 0$ or $\pi$, the fixed points of Eqs.\eqref{SymmetricTunneling} are $\bar{p}=0$ and $\cos\bar{\theta}=-\Delta/\alpha$. The oscillations of an initial population imbalance and phase difference are described by $\dot{p}=(\Delta+\alpha)\theta$, $\dot{\theta}=-(\Delta+\alpha-\gamma)p$, which results in a finite oscillation frequency $\omega=\sqrt{(\Delta+\alpha)(\Delta+\alpha-\gamma)}$. In the absence of the bare tunneling parameter $\Delta$, the frequency becomes $\sqrt{\alpha(\alpha-\gamma)}$. 

Now we discuss the time evolution of the condensates in an asymmetric double well. For $\sin\bar{\theta}=0$, the fixed points of Eq.\eqref{GeneralTunnelingEquations} are solved by
\begin{equation*}
\epsilon+(\gamma-\alpha) \bar{p}\pm\frac{\beta(1-2\bar{p}^2)-\Delta \bar{p}}{\sqrt{1-\bar{p}^2}}=0,
\end{equation*}
where the plus and minus signs correspond to $\bar{\theta}=0$ and $\pi$ respectively. For $\sin\bar{\theta}\neq 0$, the fixed points of the dynamics are determined by
\begin{equation}\label{FixedPoints}
\Delta+\beta\bar{p}+\alpha\sqrt{1-\bar{p}^2}\cos\bar{\theta}=\epsilon+\gamma\bar{p}+\beta\sqrt{1-\bar{p}^2}=0,
\end{equation}
which has the solution
\begin{equation*}
\bar{p}=\frac{\beta\Delta-\alpha\epsilon}{\alpha\gamma-\beta^2},\sqrt{1-\bar{p}^2}\cos\bar{\theta}=\frac{\beta\epsilon-\gamma\Delta}{\alpha\gamma-\beta^2}.
\end{equation*}
Specifically, for the case of $\epsilon=\Delta=0$, the fixed points are located at $\bar{p}=0$ and $\bar{\theta}=\pm\pi/2$, which implies a vanishing population imbalance and a $\pi/2$ phase difference. Near the fixed points $(\bar{p},\bar{\theta})=(0,\pm\pi/2)$, the time evolutions of the population imbalance and the relative phase are determined by
\begin{equation}
\dot{p}=\pm\beta p-\alpha\theta,\hbar\dot{\theta}=\gamma p\mp \beta\theta.
\end{equation}
Evidently, the atomic current is proportional to a linear combination of the population imbalance and the relative phase, and the relative phase is proportional to the linear combination of the population imbalance and the relative phase itself. For given $\alpha$, $\beta$ and $\gamma$, the time evolution of $p$ and $\theta$ are sinusoidal, which can be easily seen through $\ddot{p}+(\alpha\gamma-\beta^2)p=0$ and $\ddot{\theta}+(\alpha\gamma-\beta^2)\theta=0$, where $\omega=\sqrt{\alpha\gamma-\beta^2}$ is the frequency of oscillation. Hence, the time evolutions of the population imbalance and the relative phase are governed by the generalized harmonic oscillator, where the Hamiltonian is defined by $H=(\alpha \theta^2 \mp 2\beta p\theta+\gamma p^2)/2$.

\section{Magnetic system as a realization}
We can formulate the coupled mode equations for a double-well BEC into nonlinear dynamics of a classical spin ${\mathbf S}$ as
\begin{subequations}\label{Spin_Dynamics}
\begin{gather}
\dot{S}_x=-(\epsilon+\beta S_x+\gamma S_z)S_y,\\
\dot{S}_y=(\epsilon+\beta S_x+\gamma S_z)S_x-(\Delta+\alpha S_x + \beta S_z)S_z,\\
\dot{S}_z=(\Delta+\alpha S_x + \beta S_z)S_y,
\end{gather}
\end{subequations}
using the mapping $S_x=\psi_1^*\psi_2+\psi_1\psi_2^*$, $S_y=-i(\psi_1^*\psi_2-\psi_1\psi_2^*)$, $S_z=|\psi_1|^2-|\psi_2|^2$. The spin components satisfy the standard Poisson bracket $\{S_i,S_j\}=\epsilon_{ijk}S_k$ and $\Delta'=\Delta+\alpha S_x + \beta S_z$ and $\epsilon'=\epsilon+\beta S_x+\gamma S_z$ can be regarded as the effective magnetic fields in the $x$ and $z$ directions respectively. Eqs.\eqref{Spin_Dynamics} describe the anisotropic interactions of a single spin in an effective external magnetic field, which can be derived from the Hamiltonian
\begin{equation}
H=\Delta S_x+\epsilon S_z+\frac{\alpha}{2} S_x^2+\beta S_xS_z +\frac{\gamma}{2} S_z^2.
\end{equation}
Here the spin has a magnetic easy axis along the $y$-axis, $\Delta$ and $\epsilon$ represent the transverse magnetic fields, $\alpha$, $\beta$ and $\gamma$ are the second-order magnetic anisotropy parameters. If we write $\alpha=2(D+F)$ and $\gamma=2(D-F)$, the spin Hamiltonian in zero field can be written as $H=-DS_y^2+F(S_x^2-S_z^2)+\beta S_xS_z$, where $D$ represents the uniaxial anisotropy parameter and $F$ represents the transverse anisotropy parameter \cite{NanoMagnet}. For $\bar{S}_y=0$, the steady-state solutions for Eq.\eqref{Spin_Dynamics} are determined by
\begin{equation*}
(\epsilon+\beta \bar{S}_x+\gamma \bar{S}_z)\bar{S}_x-(\Delta+\alpha \bar{S}_x + \beta \bar{S}_z)\bar{S}_z=0.
\end{equation*}
For $\bar{S}_y\neq 0$, the steady-state solutions for Eqs.\eqref{Spin_Dynamics} are determined by $\epsilon+\beta \bar{S}_x+\gamma \bar{S}_z=\Delta+\alpha \bar{S}_x + \beta \bar{S}_z=0$, which are solved by
\begin{equation}
\bar{S}_x=\frac{\beta\epsilon-\gamma\Delta}{\alpha\gamma-\beta^2},
\bar{S}_z=\frac{\beta\Delta-\alpha\epsilon}{\alpha\gamma-\beta^2},\bar{S}_y=\pm \sqrt{1-\bar{S}_x^2-\bar{S}_z^2}.
\end{equation}
Specifically, for the case of zero field $(\epsilon=\Delta=0)$, the solution $(\bar{S}_x,\bar{S}_y,\bar{S}_z)=(0,\pm 1,0)$ corresponds to a spin lying along the easy axis, namely the $y$-axis. This implies the populations of the two condensates are the same and the two condensates have a $\pi/2$ phase difference. As shown in Fig.\ref{SpinDynamics}, the stability of the fixed points varies with the parameters $\epsilon$, $\Delta$, $\alpha$, $\beta$ and $\gamma$. Therefore, it is also possible to use a classical magnet with nonlinear interactions \cite{QuantumSpinGlass, QuantumMolecularMagnet} to simulate a nonlinear quantum system and the anomaly in gauge structures associated with the critical surface.

\end{appendix}

\end{document}